\documentclass[twocolumn, preprintnumbers, 
endnote,nofootinbib,prl,9pt, superscriptaddress]{revtex4}

\usepackage{epsfig,subfigure}

\usepackage{epstopdf}

\usepackage[utf8]{inputenc}
\usepackage{graphicx}
\usepackage{amssymb}
\usepackage{amsxtra}
\usepackage{amsmath}
\usepackage{booktabs,multirow,tabularx}
\usepackage{slashed}
\usepackage{float}
\usepackage{placeins}
\usepackage{rotating}
\usepackage{lscape}
\usepackage{color}
\usepackage{hyperref}

\usepackage[Q=yes,pverb-linebreak=no]{examplep}

\DeclareGraphicsExtensions{.eps}
\graphicspath{{./}}

\newcommand{\vev}[1]{\langle {#1} \rangle}
\newcommand{\lsim}{\lesssim}
\newcommand{\gsim}{\gtrsim}

\newcommand{\ord}[1]{\mathcal{O}{(#1)}}
\newcommand{\eq}[1]{Eq.~(\ref{#1})}

\def\beq{\begin{equation}}
\def\bea{\begin{eqnarray}}
\def\eeq{\end{equation}}
\def\eea{\end{eqnarray}}
\def\beqnl{\begin{align}}
\def\endal{\end{align}}
\def\abs#1{\left|#1\right|}

\definecolor{red1}{cmyk}{0,1,1,0.1}
\definecolor{blue1}{cmyk}{1,0,0,0}

\DeclareFontFamily{U}{cbgreek}{}
\DeclareFontShape{U}{cbgreek}{m}{n}{
        <-6>    grmn0500
        <6-7>   grmn0600
        <7-8>   grmn0700
        <8-9>   grmn0800
        <9-10>  grmn0900
        <10-12> grmn1000
        <12-17> grmn1200
        <17->   grmn1728
      }{}
\DeclareFontShape{U}{cbgreek}{bx}{n}{
        <-6>    grxn0500
        <6-7>   grxn0600
        <7-8>   grxn0700
        <8-9>   grxn0800
        <9-10>  grxn0900
        <10-12> grxn1000
        <12-17> grxn1200
        <17->   grxn1728
      }{}

\makeatletter
\newcommand{\normalorbold}{%
  \ifnum\pdf@strcmp{\math@version}{bold}=\z@ bx\else m\fi
}
\makeatother

\begin{document}


\title{\boldmath Galactic Dark Matter Population as the Source of Neutrino Masses}

\author{Hooman Davoudiasl\footnote{email: hooman@bnl.gov}
}

\author{Gopolang Mohlabeng\footnote{email: gmohlabeng@bnl.gov}
}

\affiliation{Physics Department, Brookhaven National Laboratory,
Upton, New York 11973, USA}

\author{Matthew Sullivan\footnote{email: mattsullivan14916@ku.edu}
}

\affiliation{Physics Department, Brookhaven National Laboratory, Upton, New York 11973, USA}

\affiliation{Department of Physics and Astronomy, University of Kansas, Lawrence, Kansas, 66045 USA}

\begin{abstract}

We propose that neutrino masses can be zero {\it in vacuo} and may be generated 
by the local distribution of dark matter through a 
feeble long range scalar force.  
We discuss potential phenomenological constraints and implications of this 
framework.  Our model typically implies that the cosmic neutrino background left over from the Big Bang 
is mostly absent in our Galactic neighborhood.  Hence, a positive detection signal from future proposed experiments, such as PTOLEMY, could in principle falsify our scenario.  
 
\end{abstract}

\maketitle

\section{Introduction\label{sec:intro}}

Of all the problems of particle physics that remain unresolved in the Standard Model (SM), the presence 
of cosmic dark matter (DM) as well as non-zero neutrino masses $m_\nu \lsim 0.1$~eV and mixing \cite{Fukuda:1998mi, Ahmed:2003kj} arguably have some of 
the strongest and most robust empirical evidence. 
However, after years of experimental and theoretical investigations, both the nature of DM and the origin of neutrino masses and mixing remain unknown.  Both neutrinos and DM, while apparently distinct in character, share the feature of having feeble - at best, in the case of DM - interactions 
with atoms and hence pose a challenge to measurements of their properties.  While neutrinos are known to have interactions other than gravitational, the same cannot be said with certainty about DM.  

The above state of affairs allows one some space for speculation about possible {\it exotic} interactions of neutrinos and DM.  While we do not know the spectrum of DM states, neutrinos are characterized by the smallest non-zero masses known in Nature.  
In fact, the minimal SM predicts that they should be massless, so their tiny sub-eV masses point to some new physics beyond the SM.  Though we accept this premise in our work, we approach it from a radically different point of view: that {\it the small but non-zero masses of neutrinos are not an inherent vacuum property, but the result of a long range scalar potential sourced by DM distributions.}  
Long range forces have received much attention in the literature due to their various implications for dark sector dynamics. The notion of a long range force was introduced by Ref. \cite{Lee:1955vk} and their possible applications to dark matter interactions have been studied from the smallest scales in our Galactic halo to the largest scales in cosmology \cite{Friedman:1991dj, Gradwohl:1992ue, Dolgov:1999gk, Farrar:2003uw, Gubser:2004du, Gubser:2004uh, Nusser:2004qu, Kesden:2006vz, Kesden:2006zb, Farrar:2006tb, Davoudiasl:2017pwe, Berlin:2016woy, Krnjaic:2017zlz, Davoudiasl:2018ltz}.

Adopting the formalism introduced in Ref. \cite{Gubser:2004du}, we consider a long range force between dark matter and neutrinos, which is mediated by a light scalar $\phi$.  For a discussion of potential theoretical motivations for such a scalar and related questions, we refer the interested reader to Ref.~\cite{Gubser:2004du}, for example. 
If dark matter sources neutrino masses then the neutrinos may be massless in empty space, but acquire small masses near non-trivial populations of DM. 
This scenario would imply that the neutrino mass matrix can vary substantially throughout space and with time. In particular, neutrinos would have very different properties in different parts of our Galaxy. 

In what follows, we will provide a simple phenomenological model of how the above neutrino mass generation mechanism can be realized.  We will then address some of the potential constraints that may apply to our scenario; it is shown, generally speaking, that the most obvious concerns about the 
viability of our idea can be addressed.  Next, we will focus on possible signals and tests of our hypothesis.  Some speculations and a summary will be presented in closing.
For possible effects of astrophysical backgrounds on neutrino properties, in a different framework, please see Ref. \cite{Fardon:2003eh}.

\section{Dynamics\label{sec:dynam}}

The basic interactions of interest for our analysis are given by 
\beq
{\cal L}_i = - g_X \phi\, \bar X X - g_\nu \,\phi\, \bar \nu\, \nu\,,
\label{Li}
\eeq
where $X$ is a DM fermion and $\nu$ is a neutrino in the SM.  Here, we assume that both particles are {\it Dirac} fermions, however our mechanism can accommodate {\it Majorana} masses for the SM neutrinos if there is a mass term for right handed neutrinos in the Lagrangian.  
In what follows, we will take the couplings of $\phi$ to other SM states to be negligible.  The above interactions can be straightforwardly generalized to include a matrix valued  $g_\nu$ that would yield the requisite mixing angles and masses.  
We will adopt $m_\nu \sim 0.1$~eV as a reasonable representative value for neutrino mass eigenvalues, where a mild variation can accommodate the current inferred mass squared differences. 
We note that current oscillation data allows for one of the mass eigenvalues to be $\ll ~0.1$ eV and in principle even zero; we will address this scenario later in the text. 
The mass terms of interest, {\it in vacuo}, are given by 
\beq
{\cal L}_m = -m_X \bar X X - \frac{1}{2} m_\phi^2\, \phi^2\,,
\label{Lm}
\eeq
where $m_X$ and $m_\phi$ are the masses of $X$ and 
$\phi$, while neutrinos are {\it massless}, in empty space. We will focus on values of $m_{X} \sim $ GeV though our conclusions do not strongly depend on this choice. In the presence of a constant background $\phi$, neutrinos have an apparent mass of 
\beq
m_\nu \equiv g_\nu \phi\,,
\label{mnu}
\eeq
which can be positive or negative. This mass term can be made positive, as is typical, by performing a chiral transformation of the neutrino field. We shall use the positive mass convention.

We assume a force between the non-relativistic dark matter and neutrinos given by a Yukawa potential of the form
\beq
V_{\phi} (r) = - \frac{g_{X} g_{\nu \,}}{4\pi ~r} e^{-m_{\phi}~r}, 
\label{potphi}
\eeq
where $r$ is the distance between the two interacting species. The force is attractive if $g_{X}$ and $g_{\nu}$ have the same sign, and is repulsive if $g_{X}$ and $g_{\nu}$ have opposite sign. We shall see later that, in order to have positive masses, the force between dark matter and neutrinos will always be repulsive in our model.

We compare the strength of the long range interaction with that of gravity. In the limit where the scalar mass $m_{\phi}$ is sufficiently small, the ratio of the Yukawa coupling to the gravitational coupling is given by 
\beq
 \beta_{f} = \frac{M_{P} g_{f}}{\sqrt{4\pi} m_{f} },
 \label{beta}
\eeq
where $M_P \approx 1.2 \times 10^{19}$~GeV is the Planck mass \cite{Patrignani:2016xqp} and fermion $f=X,\nu$.   
Given the above setup, the equation of motion for $\phi$ is given by 
\beq
(\Box + m_\phi^2) \phi = - g_X \bar X X - g_\nu \bar \nu \nu.
\label{Boxphi}
\eeq
For a fermion $f$ of number density $n_f$ and typical velocity $v_f$, Lorentz invariance 
yields 
\beq
\bar f \, f = n_f \vev{\sqrt{1 - v_f^2}}.
\label{nf}
\eeq
Here, $\vev{\dots}$ denotes an ensemble average.  For the rest of our discussion, we will only consider non-relativistic DM, with $v_X\ll 1$, well after its relic density has been established.  We will assume populations of $X$ and $\nu$ that can be considered spatially uniform and static over the distance 
and time scales relevant to our discussion, implying $\Box \phi \approx 0$ in what follows. 
  
The mean energy of neutrinos is given by $\vev{E_\nu} = m_\nu/\vev{\sqrt{1 - v_{\nu}^2}}$ and in our approximation, $\vev{E_X} \approx m_X + g_X \phi$, which ignores the kinetic energy of DM.   
Therefore, \eq{Boxphi} yields
\beq
\phi \approx  \frac{-g_X n_X}{m_\phi^2 + \omega_\nu^2}\,,
\label{phi}
\eeq
where $\omega_{\nu}^2 \equiv g_{\nu}^2 n_{\nu}/\vev{E_{\nu}}$ denotes the screening mass squared for $\phi$ induced by the neutrinos. 
Since number densities and energies are strictly positive, we note that $\phi$ and $g_X$ have opposite sign. Thus, according to \eq{mnu}, $m_\nu$ is positive if $g_X$ and $g_\nu$ have opposite sign. This confirms our statement that the Yukawa force between neutrinos and dark matter is repulsive. We will next examine how the above can allow for $m_\nu \sim 0.1$~eV from the DM distribution 
around the Solar System.  

If the screening mass $\omega^2_\nu$ from neutrinos dominates over $m_\phi^2$, then \eq{phi} reduces to 
\beq
\phi \approx \frac{-g_X n_X E_\nu}{g_\nu^2 n_\nu}\,
\label{philimit}
\eeq
If we replace $\phi$ using \eq{mnu}, we then find
\beq
\frac{m_\nu}{E_\nu} \approx \frac{-g_X n_X}{g_\nu n_\nu}\,.
\label{nurelativistic}
\eeq
which tells us that when neutrino number density is the dominant factor, neutrinos will be relativistic.
Before structure formation, the number density of neutrinos dominates DM throughout the cosmos, for the typical masses we consider here.  Until DM densities become enhanced at late times, the neutrinos would then drive $\phi\to 0$ as seen in \eq{philimit}, and the neutrinos will thus remain relativistic and nearly massless.  
Once DM clumps sufficiently it can drive out the cosmic background neutrinos from the DM dominated regions.
To see this, note that if the local DM population generates $m_\nu \sim 0.1$~eV (which acts like a repulsive potential barrier) near the Solar System the cosmic background neutrinos, characterized today by kinetic energies of $\ord{10^{-4}~\text{eV}}$, would not have enough energy to enter this region of space and would be repelled from it.

As noted above, one of the neutrinos may be much lighter than the others in our Galaxy and that species of relic neutrinos can have enough kinetic energy to enter our region of space (we will refer back to this discussion in the section on Observational Tests). 
Aside from that special case, the dominant population of neutrinos near the Earth is due to the Solar neutrino flux which yields  $n_{\nu} \sim 1 ~\rm cm^{-3}$. This number density falls off rapidly like the flux the further we move from the Sun. If one had to assume $\sim ~10^9$ stars in a kpc (as will be considered in the following analysis) neighborhood around the Earth (with $\sim$ pc distance between the stars), one can show that the average number density of neutrinos, from all the stars in this Galactic neighborhood, is $\sim 0.1 ~\rm cm^{-3}$. Then Eqs. (\ref{Boxphi}) and (\ref{nf}) imply that the stellar neutrino contribution to our potential is negligible, for $E_{\nu} \sim \rm MeV$ and our choice of parameters.

In order to keep the properties of neutrinos across the Solar System uniform, one needs to 
assume that the size of DM distribution that contributes to $m_\nu$ is much larger than ${\rm AU} \sim (10^{-18}~{\rm eV})^{-1}$.  However, in order to avoid conflict with the inferred behavior of DM on scales of $\gsim 1$~kpc, where simulations and observations seem to agree, we 
limit the range of the scalar interaction; we will adopt $m_\phi \sim 10^{-26}$~eV $\sim( 0.7 {\rm~kpc})^{-1}$ for the following discussion.
Thus one can show
\bea
m_{\nu} &\sim& 0.1 ~{\rm eV} \left( \frac{g_{\nu}}{10^{-19}}\right) \left( \frac{g_{X}/m_{X}}{10^{-19} ~{\rm GeV^{-1}}}\right) \nonumber \\
 &\times& \left( \frac{\rho_{X}}{0.3 ~{\rm GeV.cm^{-3}}}\right)  \left( \frac{10^{-26} {\rm eV}}{m_{\phi}}\right)^{2},
 \label{mnu2}
\eea
where for the above set of parameters in our local Galactic neighborhood, we find that we can ignore the screening mass in \eq{phi} (even if we include the stellar neutrino number density derived above, which is a gross overestimate on the kpc scale).  
Based on the tidal stream bounds from Refs. \cite{Kesden:2006vz, Kesden:2006zb, Carroll:2008ub} we require $\beta_{X} \lesssim 0.2$ for $m_{\phi} \lesssim 10^{-27} ~{\rm eV}$ which implies $g_{X}/m_{X} \lesssim 10^{-19} ~{\rm GeV^{-1}}$.
We are not aware of any stringent bounds on $g_{\nu}$ besides the requirement of neutrino free streaming in the early Universe (at $T \sim 1 ~{\rm eV}$)\cite{Basboll:2008fx}. Requiring that the neutrino scattering rate ($\sim g_{\nu}^{4} T$) be less than the Hubble expansion rate ($\sim T^{2}/M_P$) leads to $g_{\nu} \lesssim 10^{-7}$, which is not a severe bound in our case, given Eq.~(\ref{mnu2}).

\begin{figure}[t]
\centering
\includegraphics[scale=0.48]{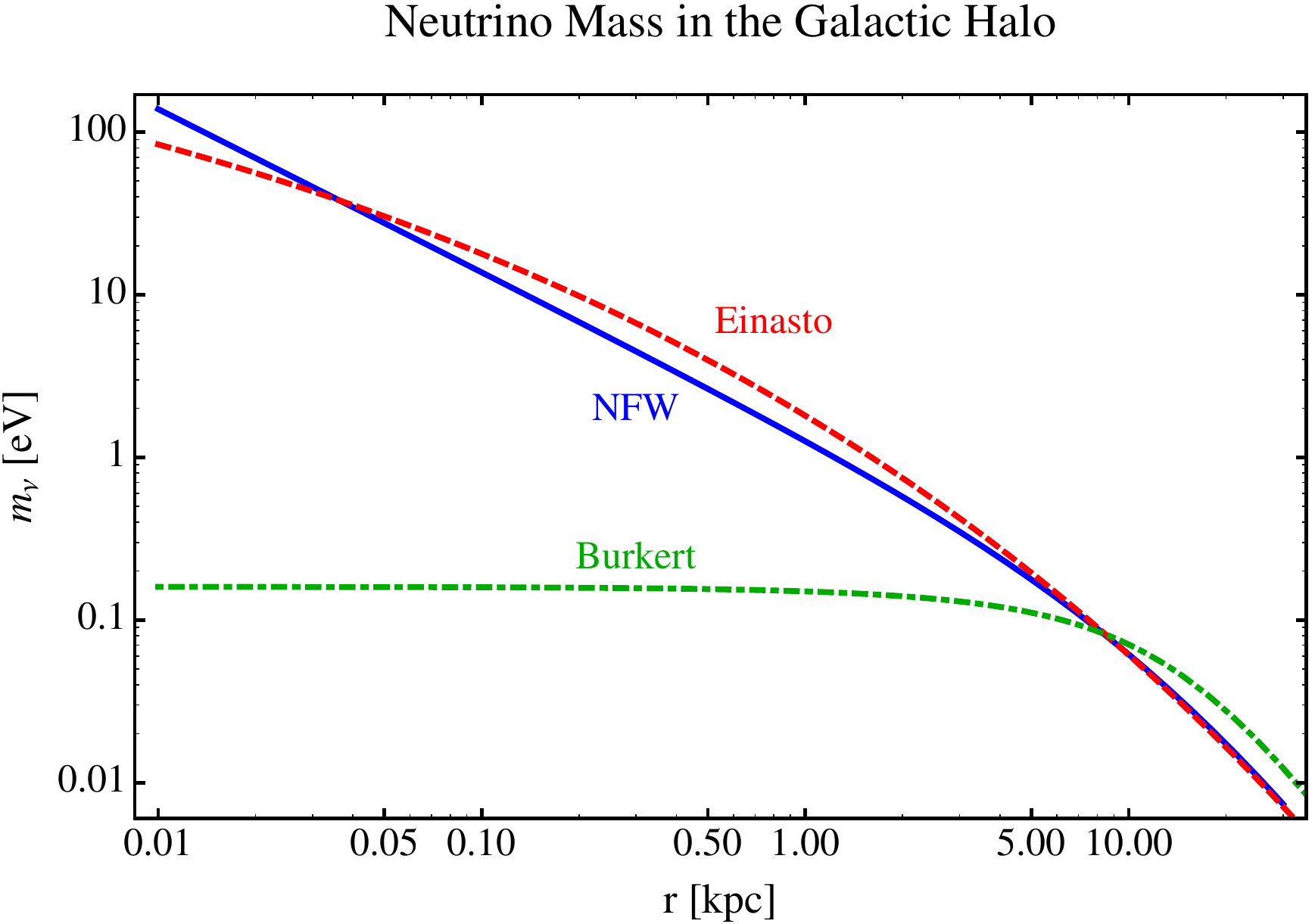}
\caption{Neutrino mass as a function of the Galactic radius for different dark matter density profiles. The red-dashed line represents a cuspy Einasto profile, the blue solid line represents a cuspy NFW profile and the green dot-dashed line represents a cored Burkert profile. For all the profiles we require that $\rho_{X} (r_{\odot}) = 0.3 {\rm~GeV.cm^{-3}}$ and $m_{X} = 0.3$ GeV, where $r_{\odot} = 8.5$ kpc.
With these parameters we obtain $m_{\nu} \sim 0.1$~eV in our local Galactic neighborhood.}
\label{numass}
\end{figure}

The above brief analysis shows that one could in principle account for neutrino masses and mixing in our Galactic neighborhood using the scalar potential sourced by DM.  To quantify this, we consider not only the dark matter distribution in our local neighborhood, but throughout the MilkyWay Galaxy. 
For illustration, we assume the three DM density profiles that are popular in the literature: the Navarro-Frenk-White (NFW), Einasto and Burkert density profiles (please see Refs.\cite{Navarro:1995iw, Bramante:2015una, Nesti:2013uwa} for their functional forms). From these, we map the distribution of the neutrino mass as a function of the Galactic radius, depicted in Fig.~\ref{numass}. 
For each profile, we require $\rho_{X} (r_{\odot}) = 0.3~{\rm GeV.cm^{-3}}$, where $r_{\odot} = 8.5$ kpc is the Galactic radius at the position of the Solar System. 
For the cuspy NFW and Einasto ($\alpha = 0.17$) profiles we assume a scale radius $R = 20$ kpc, while for the cored Burkert profile we assume a core radius $r_{c} = 16$ kpc \cite{Nesti:2013uwa, 2014JPhCS.566a2008K}.

We note here that the density of DM in other parts of the Galaxy does not get much larger than the local value near the Solar System. 
Even with an enhancement of $\ord{10^3}$, the neutrino mass $m_\nu \propto n_X$ is $\ord{100~\rm{eV}}$, which should not affect physical processes relevant to stellar and Galactic dynamics significantly. 

\subsection{Early Time Dynamics}

Let us now briefly consider earlier times, before DM has developed large scale overdensities.  In particular, let us consider 
the cosmic microwave background (CMB) era, corresponding to $T\sim 1$~eV.  At this and earlier times, we can take the DM and neutrino distributions to be homogeneous.  Also, the time scale for cosmic evolution, given by the Hubble time $H^{-1}\sim M_P/T^2$, around this era (roughly assuming radiation domination) is much larger than $m_\phi^{-1}$, and hence the potential for $\phi$ changes slowly compared to the relevant 
physical scales.  Therefore, we can use the approximation $\Box \phi \approx 0$ here.

The ratio of DM number density to entropy $s\sim g_s T^3$, with $g_s$ counting the relativistic degrees of freedom \cite{Kolb:1990vq}, is roughly given by 
\beq
\frac{n_X}{s}\sim 10^{-9}\frac{m_p}{m_X}\,,
\label{nXs}
\eeq
where $m_p$ is the proton mass (the above relation can be obtained from a similar one based on 
the baryon asymmetry).  On the other hand, the neutrino number density $n_\nu \approx T^3$ 
in the CMB era.  

Since $g_X$ and $g_\nu$ are not taken to be very different in our discussion, we then see that the neutrino scalar charge density $g_\nu n_\nu \gg g_X n_X$ at early times (for the typical masses we discuss here).  Hence, the neutrino plasma is dominant in the early Universe.  For our choice of parameters, we find $\omega_\nu \sim 10^{-19}$~eV $\gg m_\phi$, for $T\sim 1$~eV.  Thus, in the CMB era, Eqs. (\ref{philimit}) and (\ref{nurelativistic}) hold. Using \eq{nXs}, this yields
\beq
m_\nu \sim 10^{-9} \abs{\frac{g_X}{g_\nu}} \left(\frac{m_p}{m_X}\right)  
E_\nu\,. \quad \text{(CMB era)}
\label{mnuCMB}
\eeq
Therefore, for the typical range of parameters considered here, neutrinos are relativistic and 
nearly massless around the CMB era.  The above analysis is not valid at much earlier times, when the 
range of $\phi$ is limited by the horizon size instead of $\omega_\nu$.  The relevant 
temperature is given by $H\gsim \omega_\nu$, which roughly yields $T \gsim g_\nu M_P$.   
For $g_\nu \sim 10^{-19}$ we find $T\gsim 1$~GeV.  This estimate suggests that our preceding 
discussion is valid at least up to the era of Big Bang Nucleosynthesis ($T \sim 1$~MeV), which is the earliest 
cosmological time that is constrained by observations.

\subsection{Potential Constraints}
One may worry that in places where a large density of neutrinos is present considerable conflict with observations would arise.  In the current cosmological epoch, the largest neutrino number densities are those characterizing the initial stages of a supernova explosion, where 
a {\it neutrino sphere} of radius $\sim 100$~km forms, containing roughly $\ord{10^{57}}$ neutrinos.  This corresponds to an enormous number density $n_\nu^{\rm sn} \sim 10^{36}$~cm$^{-3}$.  However, these neutrinos are very relativistic, with $E_\nu \gsim 1$~MeV.  Therefore, in 
the static distribution limit, we would expect $\phi\to 0$ within the neutrino sphere and hence 
the supernova dynamics may not change appreciably. For a related discussion on long range forces acting on neutrinos in neutron stars please see Ref. \cite{Dolgov:1995hc}.
Next we show that the dark matter accumulated within the Sun would not affect the neutrino properties in the solar interior significantly. 
For the range of parameters we discuss here, dark matter would accumulate within a radius $R_{core}\sim 10^5$~km \cite{Gould:1987ir, Davoudiasl:2011fj}. Then the maximum contribution of the trapped dark matter to the neutrino mass in the Sun (near the core) would be
\beq
\delta m_{\nu} \sim \frac{g_{\nu} g_{X} N_{X}}{R_{core}},
\eeq
where $N_{X}$ is the number of dark matter particles within the core of the Sun. 
Even if one was to consider maximum dark matter accumulation due to self-interaction in the Sun, with $N_{X} \sim 10^{40}$ \cite{Kong:2014mia}, one would find 
$\delta m_{\nu} \sim 10^{-15}$~eV, which is a negligible contribution to the neutrino mass and would not have an effect on the solar neutrino dynamics.

As illustrated in Fig. \ref{numass},  as one moves away from the central parts of the Galaxy, the neutrino mass becomes smaller than $\ord{0.1 ~\rm eV}$. Given that current observational bounds on the sum of the neutrino masses, from their effects on large scale structure \cite{Allison:2015qca, LoVerde:2014rxa, Hall:2012kg}, is at or above $\ord{0.1 ~\rm eV}$, we do not expect severe constraints from these astrophysical and cosmological observations on our scenario.

At this point, we would like to address some generic model building issues.  In particular, one could ask why the neutrinos would not get masses from the Higgs mechanism, like other SM fermions.  This 
could perhaps be a consequence of underlying symmetries that forbid a neutrino-Higgs Yukawa coupling, as we will discuss next.  

For example, let us assume that right-handed neutrinos are odd under a $\mathbb Z_2$ parity, 
but none of the SM states have this parity.  As long as $\phi$ is also $\mathbb Z_2$ odd, then 
one can achieve a coupling $\phi \bar \nu_L \nu_R$ from the dimension-5 operator 
$ O_{1} = \phi H^* \bar L \nu_R/M$, where $H$ is the Higgs doublet field and $L$ is a lepton doublet, in the SM.  For $g_\nu \sim 10^{-19}$, as in the above, $\vev{H}\sim 100$~GeV implies that one then needs an effective value $M \gsim M_P$.  This suggests that the above operator is generated by very small couplings and high mass scales. If the right-handed $X$, for example, 
is $\mathbb Z_2$ odd, then one can also induce $\phi \bar X_L X_R$.  However, now, a Dirac mass term for 
$X$ cannot then be written down, if $\mathbb Z_2$ is a good symmetry.  We must then assume that  
$m_X$ is generated by a ``dark'' sector Higgs field $\Phi$ 
that spontaneously breaks $\mathbb Z_2$.   To distinguish $X$ from $\nu_R$ and ensure the stability of $X$, we will postulate that there is a $U(1)_X$ under which only $X$ is charged: $Q(X_L)=Q(X_R)=-1$.  If ${\mathbb Z}_{2}(\Phi) = -1$ one can write down $\Phi \bar X_L X_R$ and $\phi \bar X_L X_R$.  Note that The former interaction leads to a mass term for $X$ with $\vev{\Phi}\neq 0$.  For values of $m_X$ considered in our work, we may expect $\vev{\Phi} \sim 1$~GeV.   

With the above assumptions, one can write down the dimension-5 operator $O_{2} = \Phi H^* \bar L \nu_R/M$ that can contribute to $m_{\nu} \neq 0$. The effect of $O_{2}$ is negligible, with our assumptions.  To see this, note that for $\vev{\Phi} \sim ~{\rm GeV}$, $O_{2}$ would lead to a very small neutrino mass $m_\nu \lsim 10^{-8}$~eV and the long range mechanism we have introduced here would be the main source of $m_\nu \sim 0.1$~eV in and around the Solar System. 

\section{Observational Tests}
The scenario we have introduced can pose a challenge to experimental verification.  In principle, if the large scale behavior of DM shows deviations from purely gravitational dynamics, one may be led to the conclusion that there is a long range force that acts upon DM.  
The effect of this new force on neutrinos may be harder to establish.  However, as discussed earlier, our scenario typically suggests that cosmic background neutrinos do not enter the region around our Solar System at the current epoch, unless their induced mass is much less than their total kinetic energy of $\ord{10^{-4}}$ eV.
With this caveat, in the event that any of the proposed cosmic neutrino detection experiments succeeds in finding a signal, one could view this prediction of our scenario to be falsified.

The detection of cosmic background neutrinos would be a major success of the field of particle physics. To date, there are several proposed methods of detecting these neutrinos, including the Stodolsky effect \cite{Stodolsky:1974aq}, the Cavendish-like torsion balance \cite{Duda:2001hd, Strumia:2006db, Domcke:2017aqj} and interactions with Ultra-high energy cosmic rays \cite{Strumia:2006db}. 
However it seems the most promising technique for the near future is neutrino capture (please see Refs. \cite{Weinberg:1962zza, Perez-Gonzalez:2017iso} for further information) which will be exploited by the PTOLEMY experiment\cite{Betts:2013uya}, the technique of which relies on the relic neutrino having a mass not far below 0.1 eV \cite{Cocco:2017nax}. Note however that our mechanism does not allow for such relic neutrino species to be present near the Solar System. 
Hence, a near future experiment such as PTOLEMY could in principle test our model.

\section{Summary}

In this work we raise an interesting possibility that neutrinos may be massless in empty space. We introduce a model in which there exists a long range scalar mediated force between neutrinos and dark matter. 
In this scenario the background scalar potential sourced by the Galactic population of dark matter provides non-zero masses for the neutrinos. With the local dark matter densities in our Galaxy, our model can give $m_\nu \lsim 0.1$~eV neutrino masses around our solar system and different masses in other areas of the Galaxy. In addition, because this scalar potential is the source of the neutrino mass and thus determines the sign of the neutrino mass term, the force between dark matter and neutrinos will always be repulsive. As a consequence, cosmic background neutrinos have been forced out of our local Galactic neighborhood by the dark matter due to this repulsive interaction and no longer have enough energy in the present day to enter the Solar System (unless a very light mass eigenstate exists). Thus the two generic features of our proposed neutrino mass mechanism are neutrino masses which depend on local dark matter concentrations and the absence of cosmic backround neutrinos in our Galactic vicinity.

\bigskip
\bigskip

\textbf{Acknowledgements.} We would like to thank Pier Paolo Giardino, for helpful conversations on topics closely related to the current work, and Yuval Grossman, Joachim Kopp, Doug McKay and John Ralston for valuable discussions. H.D. and G.M. are supported by the United States Department of Energy under Grant Contract de-sc0012704. M.S. is supported by the United States Department of Energy grant number de-sc0017988 and by the U.S. Department of Energy, Office of Science, Office of Workforce Development for Teachers and Scientists, Office of Science Graduate Student Research (SCGSR) program. The SCGSR program is administered by the Oak Ridge Institute for Science and Education (ORISE) for the DOE. ORISE is managed by ORAU under contract number de-sc0014664.

\bibliography{Numass}
\end{document}